\documentclass[%
 reprint,
superscriptaddress,
amsmath,amssymb,
aps,
prb,
]{revtex4-2}

\usepackage{gensymb}
\usepackage{textcomp}
\usepackage{graphicx}% Include figure files
 \DeclareGraphicsExtensions{.eps,.bmp}
\usepackage{dcolumn}% Align table columns on decimal point
\usepackage{bm}% bold math
\usepackage{hyperref}% add hypertext capabilities
\begin{document}

\title{Fantastic flat bands and where to find them: The CoSn-type compounds}

\author{William R. Meier}
\affiliation{Materials Science \& Technology Division, Oak Ridge National Laboratory, Oak Ridge, Tennessee 37831}%

\author{Mao-Hua Du}
\affiliation{Materials Science \& Technology Division, Oak Ridge National Laboratory, Oak Ridge, Tennessee 37831}%

\author{Satoshi Okamoto}
\affiliation{Materials Science \& Technology Division, Oak Ridge National Laboratory, Oak Ridge, Tennessee 37831}%

\author{Narayan Mohanta}
\affiliation{Materials Science \& Technology Division, Oak Ridge National Laboratory, Oak Ridge, Tennessee 37831}%

\author{Andrew F. May}
\affiliation{Materials Science \& Technology Division, Oak Ridge National Laboratory, Oak Ridge, Tennessee 37831}%

\author{Michael A. McGuire}
\affiliation{Materials Science \& Technology Division, Oak Ridge National Laboratory, Oak Ridge, Tennessee 37831}%

\author{Craig A. Bridges}
\affiliation{Chemical Sciences Division, Oak Ridge National Laboratory, Oak Ridge, Tennessee 37831}%

\author{German D. Samolyuk}
\affiliation{Materials Science \& Technology Division, Oak Ridge National Laboratory, Oak Ridge, Tennessee 37831}%

\author{Brian C. Sales}
\affiliation{Materials Science \& Technology Division, Oak Ridge National Laboratory, Oak Ridge, Tennessee 37831}%

\date{\today}

\begin{abstract}

	Quantum interference on the kagome lattice generates electronic bands with narrow bandwidth, called flat bands. 
	Crystal structures incorporating this lattice can host strong electron correlations with non-standard ingredients, but only if these bands lie at the Fermi level. 
	In the six compounds with the CoSn structure type (FeGe, FeSn, CoSn, NiIn, RhPb, and PtTl) the transition metals form a kagome lattice. The two iron variants are robust antiferromagnets so we focus on the latter four and investigate their thermodynamic and transport properties. We consider these results and calculated band structures to locate and characterize the flat bands in these materials. We propose that CoSn and RhPb deserve the community's attention for exploring flat band physics.
\end{abstract}

This manuscript has been authored by UT-Battelle, LLC under Contract No. DE-AC05-00OR22725 with the U.S. Department of Energy. The United States Government retains and the publisher, by accepting the article for publication, acknowledges that the United States Government retains a non-exclusive, paid-up, irrevocable, world-wide license to publish or reproduce the published form of this manuscript, or allow others to do so, for United States Government purposes. The Department of Energy will provide public access to these results of federally sponsored research in accordance with the DOE Public Access Plan (http://energy.gov/downloads/doe-public-access-plan).

\maketitle

\section{Introduction}
\label{sec:Intro}

\begin{figure}
	\includegraphics[width=8.6cm]{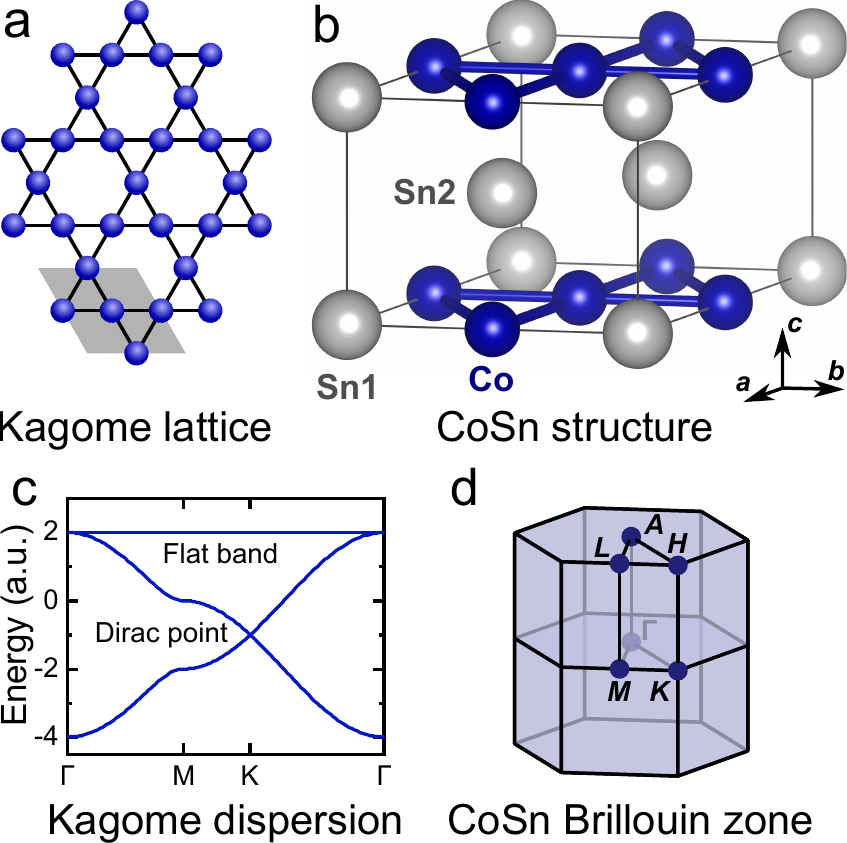}
	\caption{\label{fig:Stucture} \textbf{a}, kagome lattice with shaded CoSn unit cell. \textbf{b}, the CoSn structure with labeled sites drawn in Vesta \cite{Momma2011_Vesta3}. \textbf{c}, tight binding dispersion of kagome lattice from Ref.~\cite{Beugeling2012_TopologicalNNNHoppingIn2DLattice}. \textbf{d}, CoSn Brillouin zones with labeled high symmetry points.
	}
\end{figure}

The behavior of electrons in solids is strongly determined by the constituent atoms and the lattices on which they are arranged \cite{Kittel2004_SolidStatePhysics,Grosso2014_SolidStatePhysics}. In some lattices (e.g.~dice, Lieb, kagome), destructive quantum interference of different hopping paths generates localized electronic states in otherwise itinerant systems \cite{Leykam2018_ArtificialFlatBandSystems,Sutherland1986_LocalizationOfElectronicWaveFunctionsDueToLocalTopology}. The resulting energy bands have narrow bandwidths and appear as nearly horizontal dispersion plots, so-called flat bands (Fig.~\ref{fig:Stucture} c).

The interaction of relatively localized states drives some of the most curious phenomena in condensed matter systems; magnetism, unconventional superconductivity, and heavy-fermion physics \cite{Grosso2014_SolidStatePhysics}. Flat bands can be built with elements not usually associated with magnetism or localized states because the localization arises from the connectivity of the lattice. This provides platforms to explore strongly correlated physics with unorthodox ingredients if these flat bands are at the Fermi level. An illustrative example is superconductivity and correlated ground states in twisted bilayer graphene which arise from flat bands \cite{Cao2018_CorrelatedInsulatorMagicAngleTwistedBilayerGraphene,Cao2018_UnconventionalSuperconductivityMagicAngleTwistedBilayerGraphene}. 

In real materials, the localized character of flat bands is modified by other couplings. Notably, the spin orbit coupling can result in non-trivial band topology \cite{Sun2011_NearlyFlatbandsWithNontrivialTopology}.

The kagome lattice (Fig.~\ref{fig:Stucture} a) can generate flat bands (Fig.~\ref{fig:Stucture} c) due to hopping interference \cite{Kang2019_DiracAndFlatBandsFeSn,Kang2020_TopologicalFlatBandsCoSnARPES,Liu2020_DiracFermionsFlatBandsCoSn_ARPES}. It is composed of a pretty arrangement of hexagons and triangles depicted in Fig.~\ref{fig:Stucture}~a. This lattice can be found in Fe$_3$Sn$_2$, Co$_3$Sn$_2$S$_2$, and compounds with the HfFe$_6$Ge$_6$ or CoSn structures. 

CoSn adopts a hexagonal B35 structure type depicted in Fig.~\ref{fig:Stucture} b with space group $P6$/$mmm$ (No.~191). It is composed of a kagome lattice of Co atoms (Wyckoff site $3f$) with a Sn1 (site $1a$) centered in the hexagons. These layers are stacked along the $c$-axis and separated by honeycomb layers of Sn2 on site $2d$. In CoSn, cobalt atoms generate multiple copies of the flat band states, each associated with different $d$-orbitals \cite{Liu2020_DiracFermionsFlatBandsCoSn_ARPES}.

In total, six intermetallic compounds have been reported with the CoSn structure type \cite{Villars2012_PearsonCrystalDatabase,Bergerhoff1987_ICSD}: hexagonal-FeGe \cite{Richardson1967_EquilibriumDiagram_FeGe_System}, FeSn \cite{Nial1945_XrayStudiesOfBinaryAlloysOfTin}, CoSn \cite{Nial1945_XrayStudiesOfBinaryAlloysOfTin}, NiIn \cite{Hellner1947_InGaCompoundsWithNiPdPtCuAgAu}, RhPb \cite{ElBoragy1972_Rh-Pb}, and PtTl\cite{Zintl1935_StructPlatinumThallium}. In all but PtTl, the transition-metal is reported to take the Co site forming the kagome lattice with $p$-block element occupying the $1a$ and $2d$ sites. The site assignment in PtTl has not been determined \cite{Zintl1935_StructPlatinumThallium}, but in section~\ref{sec:Discussion_PtTl_Order} we argue this compound is ordered and Pt occupies the $3f$ Wycoff site.

We divide these metals into two groups based on valence electron count. The antiferromagnets FeGe and FeSn are composed of Fe (group 8) and a group 14 element suggesting 22 valence electrons per formula unit (F.U.). The other four compounds have one additional electron, totaling 23. CoSn and RhPb have elements from groups 9 and 14 while NiIn and PtTl are pairs from groups 10 and 13. For simplicity, we will refer to these two sub-families as the 22 and 23 electron compounds, respectively.

The antiferromagnetism of FeGe and FeSn has been thoroughly examined \cite{Beckman1972_hexFeGe_MagSusceptAndOrdering,Beckman1973_LocalVsItinerantInHexFeGe,Bernhard1984_LowTempAFM_FeGe,Bernhard1988_MagneticPhaseDiagramFeGe,Forsyth1978_LowTempAFM_hexFeGe,Haeggstroem1975_Mossbauer_hexFeGe,Stenstroem1972_ResistivityHexFeGe,Yamamoto1966_MossbauerFe5Sn3+FeSn,Yamaguchi1967_FeSnNeutronDiffraction,Sales2019_2DMagnetismFeSn,Meier2019_AFM_ReorientationCoDopedFeSn,Ligenza1972_SnMossbauerFeSn,Ligenza1971_SpinFlipFeSn,Haggstrom1975_MossbauerFeSn,Yamaguchi1967_FeSnNeutronDiffraction,Haggstrom1974_MossbauerFeGe_FeSn_CoSn}. A series of recent angular resolved photo-emission spectroscopy (ARPES) studies have investigated the electronic bands arising from the kagome lattice in FeSn \cite{Kang2019_DiracAndFlatBandsFeSn,Lin2019_DiracInFeSn} and CoSn \cite{Kang2020_TopologicalFlatBandsCoSnARPES,Liu2020_DiracFermionsFlatBandsCoSn_ARPES}. The remaining three compounds are poorly explored and the character of their flat bands is unknown.

In this family of compounds, we have $3d$, $4d$, and $5d$ transition-metals represented. We have an opportunity to examine trends across the periodic table with increasing period and spin orbit coupling strength. In FeGe and FeSn, the Fermi level lies within the flat bands and they reduce degeneracy by becoming antiferromagnets. Electron doping FeSn by swapping Co for Fe systematically suppresses magnetic order \cite{Meier2019_AFM_ReorientationCoDopedFeSn,DjegaMariadassou1969_MSn2+MSn}. Rh and Pt are rarely magnetic. If the Fermi level can be tuned to the flat bands in RhPb and PtTl by hole doping, how will the systems reduce degeneracy? Will we find magnetism, charge density waves, or superconductivity? 

In this paper we want to lay the groundwork for exploring flat band physics in the CoSn-type kagome metals. To this end, we have two main goals. First, to present and compare the thermodynamic and electrical transport properties of CoSn-type compounds. We focus on the four 23 electron compounds; CoSn, NiIn, RhPb and PtTl. Second, to discuss the energy and character of the flat bands in these kagome metals in light of our data and density functional theory (DFT) results.

In section~\ref{sec:Methods_growth}, we recount our procedures for synthesizing the six compounds and we describe the products in section~\ref{sec:Results_Products}. Next, we present the low-temperature lattice parameters, thermal expansion, heat capacity, magnetic susceptibility, and resistivity of these metals in sections~\ref{sec:Results_ThermalExpansion} through \ref{sec:Results_Resistivity}. We do not observe any phase transitions in CoSn, NiIn, RhPb or PtTl down to 1.9\,K. In the calculated band structures of these four (section~\ref{sec:Results_BandStructure}), we note that CoSn has flat bands closest to the Fermi energy ($\varepsilon_\mathrm{Fermi}$). This is supported by CoSn’s unusual magnetic susceptibility which we model as temperature dependence of Pauli paramagnetism resulting from flat bands about 20\,meV below $\varepsilon_\mathrm{Fermi}$ (section~\ref{sec:Discussion_SusceptibilityModel}). Finally, we compare the relative attractiveness of these compounds as platforms for exploring flat band physics and provide broader guidelines for other systems in section~\ref{sec:Discussion_FlatBands}.

\section{Methods}
\label{sec:Methods}

\subsection{Growth}
\label{sec:Methods_growth}

Single crystals of CoSn, FeSn, RhPb and PtTl were grown from molten metal fluxes. CoSn and FeSn were both grown from liquid Sn, RhPb from Pb and PtTl from Tl. Polycrystalline FeGe and NiIn was synthesized by arc melting the constituent elements and annealing. 

\textbf{FeGe} There are three polymorphs of FeGe with increasing temperature; hexagonal with the B35 CoSn structure, cubic with a B20 FeSi structure, and monoclinic \cite{Richardson1967_EquilibriumDiagram_FeGe_System}. Crystals of each phase can be obtained by chemical vapor transport. We were unable to obtain the hexagonal phase and chose to create the hexagonal phase by solid state reaction instead.

Equimolar iron  granules (Alfa Aesar 99.98\%) and germanium pieces (Alfa Aesar Puratronic 99.9999+\%) totaling 2\,g were arc melted together in argon. The iron pieces were placed on top of the Ge to prevent the latter shattering due to thermal shock in the arc. The button was flipped and remelted 3 times. Powder x-ray diffraction on this material showed a mix of FeGe$_2$ and Fe$_{6.5}$Ge$_4$.

Pieces of the arc melted button were annealed on alumina in an argon filled, fused-silica ampoule. This assembly was annealed at 700\textdegree C for 112\,h then quenched into water. X-ray diffraction revealed that it was composed of mostly hexagonal FeGe with about 5\,wt.\% Fe$_6$Ge$_5$.

\textbf{FeSn}
Crystals of FeSn were grown from Sn flux as in Refs.~\onlinecite{Sales2019_2DMagnetismFeSn,Meier2019_AFM_ReorientationCoDopedFeSn}. Iron granules (Alfa Aesar 99.98\%) and tin shot (Alfa Aesar, Puratronic 99.9999\%) in an atomic ratio of Fe\,:\,Sn = 2\,:\,98 were loaded into a 10\,mL alumina crucible and capped with a second upside-down crucible loosely filled with silica wool to catch the crystals. These crucibles were expertly sealed in a fused-silica ampoule under vacuum by the ORNL glass shop. The sealed silica ampoule was heated in a box furnace to 1100\textdegree C at 120\textdegree C/h and held for 12\,h, then quickly cooled at 6\textdegree C/h to 1000\textdegree C and held for 48\,h. During this hold, the furnace was opened and the ampoule was shaken with tongs to mix the melt \cite{Sales2019_2DMagnetismFeSn}. Then the furnace is cooled quickly at 6\textdegree C/h to 800\textdegree C and held for 1\,h. Finally we slowly cool to about 600\textdegree C at 1\textdegree C/h to grow the crystals. The ampoule was removed from the hot furnace, placed upside down in a centrifuge and spun. This ``spinning" process is intended to fling the liquid flux off the crystals and essentially quenches the ampoule assembly in air.

\textbf{CoSn} 
Crystals of CoSn were prepared in a 31\,g batch with cobalt pieces (Alfa Aesar 99.95\%) and tin shot (Alfa Aesar Puratronic 99.9999\%). An atomic ratio of Co\,:\,Sn = 8\,:\,92 was assembled in the same 10\,mL alumina crucible setup as FeSn. The ampoule assembly was heated in a box furnace to 1130\textdegree C at 120\textdegree C/h, held for 24\,h, then cooled at 6\textdegree C/h to 900\textdegree C and held for 1\,h. Then the furnace was slowly cooled to 618\textdegree C at 1\textdegree C/h. The ampoule was then removed and centrifuged to separate the liquid from the crystals.

\textbf{NiIn} 
The reported Ni-In phase diagrams suggest NiIn cannot be precipitated from a Ni-In liquid \cite{Ni-In_ASM1600357,Richter1998_In-Ni-Sb}. We obtained polycrystalline material by arc melting equiatomic nickel and indium metal (totaling 1.9\,g) together then annealing. Indium and its alloys can stick to the copper hearth plate of the arc melter. To mitigate this, Ni slugs (Alfa Aesar 99.995\%) and In shot (Alfa Aesar 99.9995\%) were wrapped in a piece of nickel foil (Alfa Aesar 99.95\%) to prevent direct contact between the indium and Cu hearth-plate. This procedure worked well but arc power needed to be kept low to reduce the vaporization of indium evident by its dark-blue arc (which gave indium its name\cite{lide2005_CRCHandbook_ChTheElements}). The pellet was melted and flipped 5 times. 

Powder X-ray diffraction revealed the product was mostly NiIn with minor Ni$_{2}$In$_{5}$ and Ni$_{13}$In$_{9}$. Annealing in an argon-filled fused-silica ampoule at 725\textdegree C for 134\,h removed any trace of the diffraction peaks from these impurities.

\textbf{RhPb} 
Single crystals of RhPb were grown from a lead flux in a 5.4\,g batch. Rhodium sponge (Alfa Aesar 99.95\%) and lead slugs (Alfa Aesar Puratronic 99.999\%) were placed in a 2\,mL alumina Canfield Crucible Set (CCS) \cite{Canfield2016_CanfieldCrucibleSet} with an atomic ratio Rh\,:\,Pb = 1\,:\,2 based on the binary phase diagram \cite{ElBoragy1972_Rh-Pb,Pb-Rh_ASM901860}. The crucibles were sealed in a fused silica ampoule under vacuum using a oxygen-hydrogen torch. The assembly was heated in a box furnace over 6\,h to 1000\textdegree C or 1100\textdegree C and held for 2\,h, then quickly cooled to 900\textdegree C over 4\,h. To grow the crystals, the furnace was cooled slowly to 750\textdegree C over 320\,h. The remaining Pb-rich liquid was separated from the crystals by spinning in a centrifuge.

\textbf{PtTl}
Single crystals of PtTl were obtained from a melt with an atomic ratio of Pt\,:\,Tl = 1\,:\,3 based on existing phase diagrams \cite{Pt-Tl_ASM901955,Bhan1968_Pd-Tl_Pt-Tl}. Inside a He glove box, pieces of Pt sheet (Alfa Aesar 99.9\%) and chunks of Tl (Alfa Aesar 99.999\%) totaling 6.6\,g were loaded into a 2\,mL alumina CCS. The crucibles were assembled in a fused silica tube and a valve for our gas manifold was attached and closed in the helium atmosphere. This procedure reduced the exposure of thallium metal to air while sealing the tube with a torch.

Our first reaction of Pt and Tl had a disturbing result. We heated the ampoule to 1050\textdegree C for a 24\,h hold to dissolve the platinum. When the assembly was removed from the furnace to spin in the centrifuge the inner surface of silica ampoule had darkened dramatically. We assume this is due to a reaction with the thallium vapor and could cause the ampoule to break and release the poisonous vapor. 

To reduce the risk associated with thallium weakening the silica ampoule, we urge the reader not to exceed 1000\textdegree C and to use the following furnace schedule that worked just as well. In addition, the furnace should be placed in a fume hood. The furnace was heated at 120\textdegree C/hr to 1000\textdegree C for a 24\,h hold, quickly cooled at 6\textdegree C/h to 900\textdegree C followed by a 1\,h hold. To grow crystals, the furnace was slowly cooled toward 500 \textdegree C at 2 \textdegree C/h. At 527 \textdegree the ampoule was removed and spun in a centrifuge to fling the thallium-rich melt from the crystals.

\subsection{Measurements}
\label{sec:Methods_Measurements}

Powder x-ray diffraction measurements were preformed on a PANalytical X'pert Pro diffractometer with a Cu K$\alpha$ tube and an incident beam monochrometer. Low-temperature diffraction measurements were done in an Oxford PheniX closed-cycle helium cryostat. Lattice parameters were determined from fits of powder x-ray diffraction data with HighScore Plus. The high temperature diffraction data were collected using an Anton Paar XRK900 reactor stage on an PANalytical Empyrian diffractometer employing Cu K$\alpha$ radiation under flowing UHP He gas. The stage height was initially aligned using computer controlled height adjustment, and the height was subsequently auto-corrected for thermal expansion of the Macor sample holder during the measurement.

Heat capacity of each sample was measured with Apiezon N-grease using the heat capacity option of a Quantum Design Physical Property Measurement System (PPMS) or DynaCool PPMS. Magnetization measurements were performed at 10\,kOe in a Quantum Design Magnetic Property Measurement System (MPMS) in plastic drinking straws. Resistivity vs temperature measurements were also performed in the PPMS using the ac-transport option. Both single crystals and polycrystalline samples were polished into bars and 4 platinum wires were attached with silver epoxy (EPO-TEK H20E).

\subsection{Density functional theory}
\label{sec:Methods_DFT}

Electronic band structures and density of states were calculated for CoSn, NiIn, RhPb, and PtTl based on density functional theory (DFT) with Perdew-Burke-Ernzerhof exchange-correlation functional \cite{Perdew1996_GGA}, as implemented in the VASP code \cite{Kresse1996_AbInitioEnergyCalculations}. Spin orbit coupling was included in these calculations self-consistently. The interaction between ions and electrons was described by the projector augmented wave method \cite{Kresse1999_ProjectorAugmentedWaveMethod}. The kinetic energy cutoffs for the plane-wave basis are 350\,eV for CoSn, NiIn, and RhPb and 308\,eV for PtTl.  12$\times$12$\times$12 and 20$\times$20$\times$20 $k$-point meshes were used for structural optimization and density of states calculations, respectively. The optimized lattice parameters reported in table~\ref{tab:LatticeSummary} compare well with the measured lattice parameters at 15\,K.

%The optimized lattice parameters are $a = 5.290$\,\AA, c = 4.224 \AA (CoSn), a = 5.276 \AA, c = 4.376 \AA (NiIn), a = 5.731 \AA, c = 4.505 \AA (RhPb), and a = 5.698 \AA, c = 4.787 \AA (PtTl), which are in good agreement with experimental values in Table~\ref{tab:LatticeSummary}.

\section{Results}
\label{sec:Results}

\subsection{Products}
\label{sec:Results_Products}

\begin{figure}
	\includegraphics[width=8.6cm]{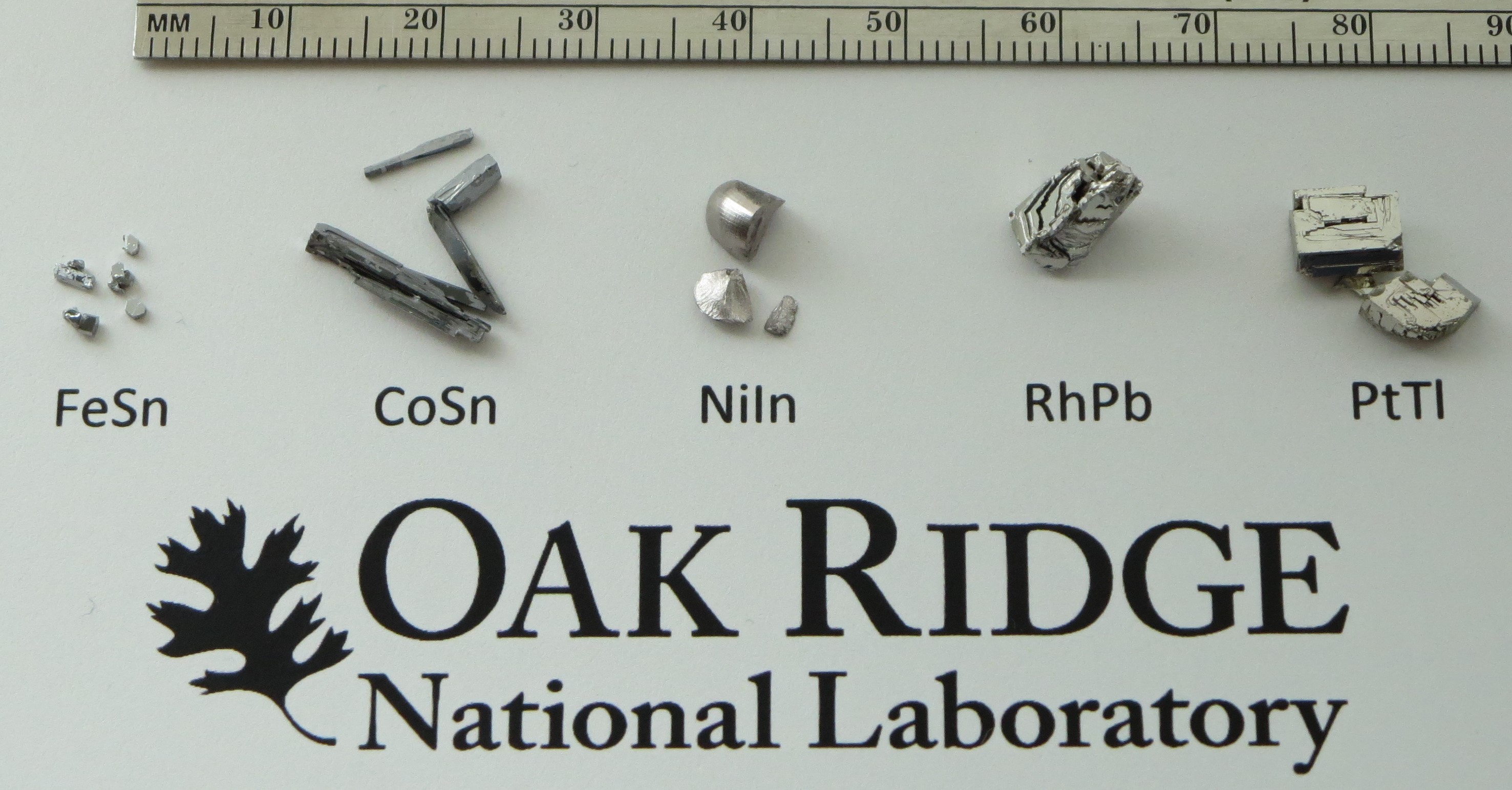}
	\caption{\label{fig:GroupPhoto} %\textbf{Will take new photo with FeGe and smaller logo.} 
		A photo of the CoSn-type samples with steel scale. The pale-blue metallic luster of CoSn is evident but the pale-pink hue of NiIn doesn't show up well.
	}
\end{figure}

The crystals of FeSn, CoSn, RhPb and PtTl obtained by flux growth (Fig.~\ref{fig:GroupPhoto}) were generally blocky, hexagonal prisms often with mirror-like faces and a bright metallic luster. Crystals of CoSn tended to be more elongated and had a subtle blue-gray hue. The annealed polycrystalline sample of FeGe had a metallic luster and a pale gray color. Notably, the annealed polycrystalline NiIn was metallic with an unusual pale pink color. This doesn't appear to be a surface film as it is still present on freshly broken surfaces and the ground powder had a distinct violet tint. Both NiIn and PtTl are in a group of intermetallics expected to be colored like red PdIn and purple AuAl$_2$\cite{Steinemann1997_ColorInPettiforsStructureMaps,Steinemann2002_IntermetallicsColorAndOpticalProperties}.

All compounds are stable in air for months. Crystals of FeSn, CoSn, RhPb, and PtTl exhibit good (001) cleavage.

\subsection{Thermal Expansion}
\label{sec:Results_ThermalExpansion}

\begin{figure}
	\includegraphics{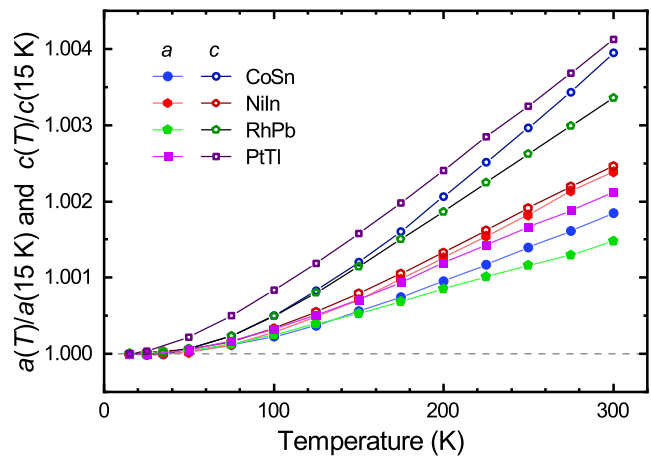}
	\caption{\label{fig:ThermalExpansion} Thermal expansion of 23 electron kagome metals. The lattice parameters are normalized by dividing by the 15\,K value of each compound. These four compounds show larger thermal expansion along the $c$-axis than $a$ but NiIn is nearly isotropic.
	}
\end{figure}

\begin{figure}
	\includegraphics{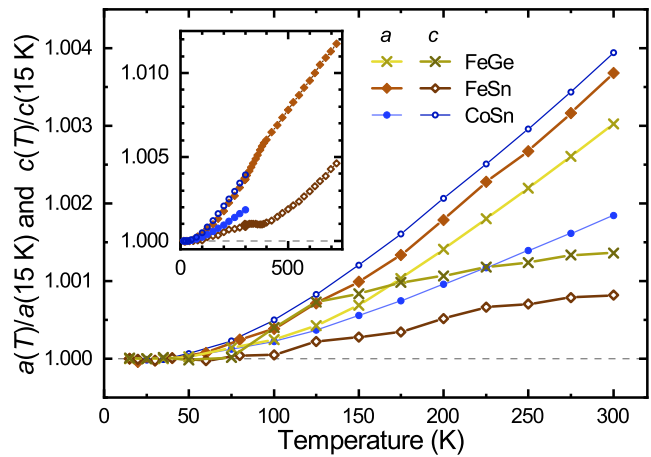}
	\caption{\label{fig:FeSn_ThermalExpansion} The thermal expansion of FeGe, FeSn, and CoSn. Note that FeGe and FeSn show the opposite thermal expansion anisotropy to CoSn (and the other compounds without magnetic order). The moment reorientation in FeGe generates feature around 100\,K. The inset shows the thermal expansion of FeSn up to 725\,K. Antiferromagnetism appears to suppress $c$-axis expansion below $T_\mathrm{N}$.}
\end{figure}

\begin{table*}
	\caption{Lattice parameters of CoSn-type compounds at 15 and 300\,K. The relaxed lattice parameters from DFT are presented for those without magnetic order. The thermal expansion coefficient, $\alpha$, between 200 and 300K is presented for each principle direction.
		\label{tab:LatticeSummary}}
	\begin{tabular}{l|c|c|c||c|c|c||c|c}
		 &$a$ DFT &$a$(15\,K) &$a$(300\,K) &$c$ DFT &$c$(15\,K) &$c$(300\,K) &$\alpha_{a}$ &$\alpha_{c}$\\
		&(\AA) &(\AA) &(\AA) &(\AA) &(\AA) &(\AA) &(10$^{-6}$/K) &(10$^{-6}$/K)\\
		\hline
		FeGe & &4.98515(10) &5.00025(10) & &4.04817(9) &4.04817(9) &16.1(1) &3.0(3)\\
		FeSn & &5.27651(11) &5.29591(11) & &4.44430(8) &4.44795(8) &18.6(4) &3.0(5)\\
		CoSn &5.290 &5.26926(6) &5.27896(6) &4.224 &4.24310(5) &4.25985(5) &8.83(4) &18.7(3)\\
		NiIn &5.276 &5.22955(9) &5.24203(8) &4.376 &4.33892(7) &4.34962(7) &11.33(18) &11.38(9)\\
		RhPb &5.731 &5.66601(20) &5.67438(20) &4.505 &4.41267(10) &4.42746(10) &6.15(15) &14.83(8)\\
		PtTl &5.698 &5.60172(5) &5.61359(5) &4.787 &4.62755(3) &4.64664(3) &9.17(7) &17.04(16)\\
		
	\end{tabular}
\end{table*}

The lattice parameters of the kagome metals are reported in table~\ref{tab:LatticeSummary} and the thermal expansion of the 23 electron compounds are presented in Fig.~\ref{fig:ThermalExpansion}. NiIn shows nearly isotropic relative expansion, but the other three show notably faster expansion along the $c$-axis. The room temperature thermal expansion coefficients, $\alpha$, in table~\ref{tab:LatticeSummary} were calculated using the slope of a linear fit of $a(T)$ and $c(T)$ between 200 and 300\,K.

Figure~\ref{fig:FeSn_ThermalExpansion} emphasizes how antiferromagnetism leads to strikingly different lattice evolution in the 22 electron metals FeGe and FeSn. In contrast to the 23 electron metals, these compounds show smaller expansion along the $c$-axis than $a$. Strong magnetic coupling between the kagome sheets of Fe \cite{Sales2019_2DMagnetismFeSn} seems to resist thermal expansion along the $c$-axis. The inset shows that the thermal $c$-axis expansion of FeSn is suppressed until magnetic order melts at the N\'eel temperature evident as change in slope near 365\,K. A moment reorientation near 100\,K appears as an anomaly in the FeGe data.

\subsection{Heat Capacity}
\label{sec:Results_HeatCapacity}

\begin{table*}
	\caption{Summary of measured physical parameters of the CoSn-type compounds including the N\'eel temperature, $T_\mathrm{N}$, of the two antiferromagnets. The Sommerfeld coefficient, $\gamma$, from and Debye temperatures, $\theta_\textrm{D}$ are derived from heat capacity measurements. $\chi_\mathrm{Core}$ is the diamagnetic contribution from the noble gas cores of each atom from Kittel \cite{Kittel2004_SolidStatePhysics}. Next, the powder averaged susceptibility ($\chi_\mathrm{pow} = (2 \chi_{\perp c} + \chi_{\parallel c})/3$) and the magnitude of magnetic anisotropy ($\Delta\chi = \chi_{\perp c} - \chi_{\parallel c}$) from the compounds without magnetic order at 50\,K. Finally the anisotropic residual resistivity ($\rho(2\,\mathrm{K})$) of the compounds. A single value is reported for the polycrystalline FeGe and NiIn samples.
		\label{tab:PropertySummary}}
	\begin{tabular}{l|c|c|c|c|c|c|c|c}
		 &$T_\mathrm{N}$ &$\gamma$ &$\theta_\textrm{D}$ &$\chi_\mathrm{Core}$ &$\chi_\mathrm{powder}$ &$\Delta\chi$ &$\rho_{\parallel c}$(2\,K) &$\rho_{\perp c}$(2\,K)\\
		&(K) &(mJ/mol\,f.u.\,K$^2$) &(K) &\multicolumn{3}{c|}{(cm$^3$/mol\,f.u.)} &\multicolumn{2}{c}{(\textmu $\Omega$\,cm)}\\
		\hline
		FeGe &411\cite{Stenstroem1972_ResistivityHexFeGe} &7.6(3) &427(10) &-0.000039 &0.004 &  &\multicolumn{2}{c}{23.35(2)}\\
		FeSn &365\cite{Sales2019_2DMagnetismFeSn} &10.98(5) &314(3) &-0.000047 & &  &1.01(6) &1.65(3)\\
		CoSn &  & 3.68(4) &380(5) &-0.000047 &0.000073 &0.000081 &0.651(3) &10.81(5)\\
		NiIn &  &3.98(1) &457(4) &-0.000047 &-0.000007 &  &\multicolumn{2}{c}{0.70(2)}\\
		RhPb &  &1.36(15) &251(4) &-0.000071 &-0.000063 &0.000100 &2.30(2) &1.35(1)\\
		PtTl &  &1.60(10) &195(1) &-0.000086 &-0.000074 &0.000053 &0.23(7) &0.81(1)\\
		
	\end{tabular}
\end{table*}

\begin{figure}
	\includegraphics{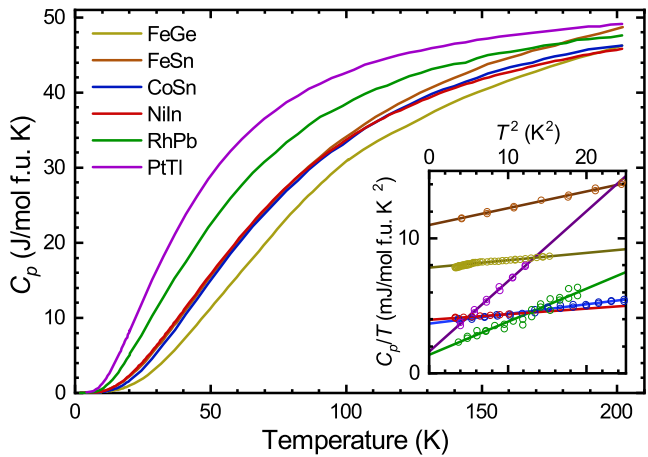}
	\caption{\label{fig:HeatCapacity} Heat capacity of the 6 kagome metals. The inset presents low-temperate fits used to estimate the Sommerfeld parameter, $\gamma$, and estimate the Debye temperature ($\theta_\textrm{D}$) presented in table~\ref{tab:PropertySummary}. The moment reorientation transition in FeGe is evident as a change in slope around 100\,K.}
\end{figure}

The heat capacity vs temperature of the CoSn-type compounds is depicted in Fig.~\ref{fig:HeatCapacity}. The curves shift to lower temperatures with increasing molar mass from FeGe $\rightarrow$ (FeSn, CoSn, and NiIn) $\rightarrow$ RhPb $\rightarrow$ PtTl, reflecting a decrease in the Debye temperature, $\theta_\textrm{D}$ (Table~\ref{tab:PropertySummary}). 

The inset in this figure shows low-temperature fits used to estimate the Sommerfeld coefficient, $\gamma$. These values are less certain for FeGe due to a subtle feature around 2.5\,K. We attribute this to a minor second phase observed in the polycrystalline sample. Despite this, the antiferromagnets FeGe and FeSn have markedly larger $\gamma$'s than compounds without order. 

The values of $\gamma$ we obtain for FeGe, FeSn, and CoSn are comparable to those estimated from low-temperature heat capacity in Ref.~\onlinecite{Larsson1974_SpecificHeat_FeGe_FeSn_CoSn} (8.46(2), 10.95(1), and 4.57(3) mJ/mol\,f.u.\,K$^2$ respectively). They obtained different estimates of $\theta_\textrm{D}$ (341, 303, and 299\,K respectively) from their more detailed analysis. 

FeGe and FeSn are both antiferromagnetic across the entire measured temperature range\cite{Beckman1973_LocalVsItinerantInHexFeGe,Beckman1972_hexFeGe_MagSusceptAndOrdering,Stenstroem1972_ResistivityHexFeGe,Bernhard1988_MagneticPhaseDiagramFeGe,Sales2019_2DMagnetismFeSn,Kakihana2019_ElectronicStatesFeSn+CoSn,Yamaguchi1967_FeSnNeutronDiffraction}. The moment reorientation of FeGe is evident as a change in slope of $C_p(T)$ around 100\,K \cite{Beckman1973_LocalVsItinerantInHexFeGe,Beckman1972_hexFeGe_MagSusceptAndOrdering,Bernhard1988_MagneticPhaseDiagramFeGe}. %In addition the magnetic contribution to heat capacity of FeGe and FeSn is evident in the steeper slopes of $C_p(T)$ at 200\,K relative to the the other 4 compounds.

\subsection{Magnetic susceptibility}
\label{sec:Results_MagneticSusceptibility}

\begin{figure}
	\includegraphics{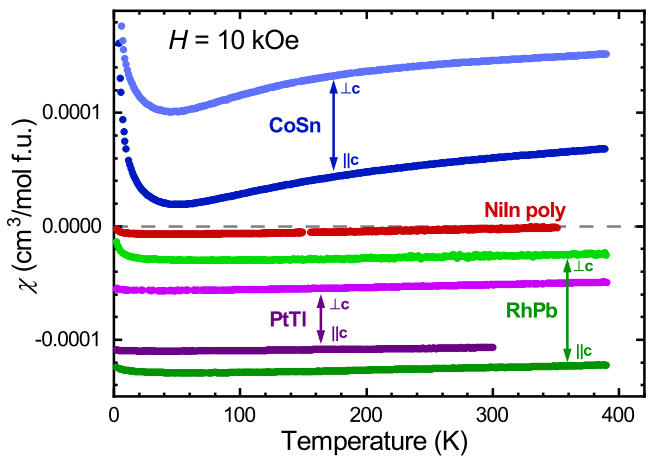}
	\caption{\label{fig:ChiT} Anisotropic magnetic susceptibility of the 4 kagome metals without magnetic order. The data for NiIn is from the annealed polycrystalline powder. The other 3 show a lower susceptibility with field along the $c$-axis. CoSn shows an pronounced increase in $\chi$ with temperature.}
\end{figure}

\begin{figure}
	\includegraphics{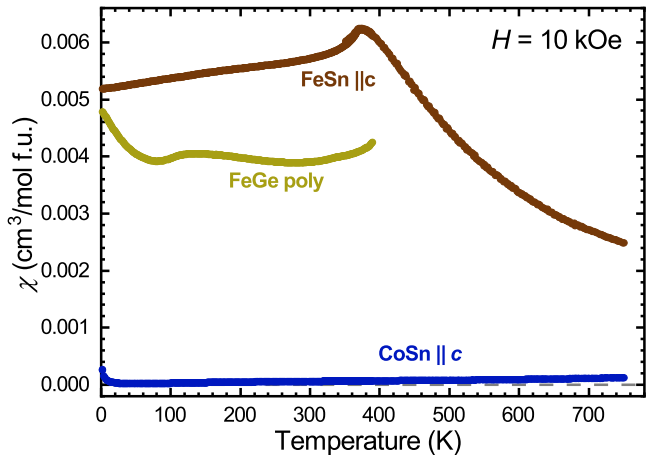}
	\caption{\label{fig:ChiT+Mag} FeGe and FeSn show a dramatically larger magnetic response than CoSn and the other metals without magnetic order. The FeGe data was obtained with an annealed polycrystalline sample.}
\end{figure}

The magnetic susceptibility data for the CoSn-type compounds held some surprises for us. We anticipated relatively isotropic and temperature-independent susceptibility for the 23 electron materials. Figure~\ref{fig:ChiT} reveals the situation is more complicated. Polycrystalline NiIn shows the smallest response with weak temperature dependence. RhPb and PtTl are diamagnetic and show only a subtly rising $\chi(T)$ apart from a minor Curie contribution at low temperatures. CoSn shows a stronger Curie tail and an unusual rising susceptibility contribution above 50\,K which is nearly identical parallel and perpendicular to $c$-axis. Finally, we observe an essentially temperature-independent difference between the $c$-axis and in-plane susceptibility of CoSn, RhPb, and PtTl (see Table~\ref{tab:PropertySummary}). In all three cases the susceptibility is smaller with $\bm{H}\parallel\bm{c}$ and the difference is the same magnitude as the powder averaged susceptibility.

Figure~\ref{fig:ChiT+Mag} illustrates the dramatically weaker magnetic response of CoSn relative to the antiferromagnets FeGe and FeSn. This relatively small susceptibility suggests that magnetic order has no role in the magnetic response of CoSn, NiIn, RhPb or PtTl. The anisotropic susceptibility of FeGe \cite{Beckman1973_LocalVsItinerantInHexFeGe,Beckman1972_hexFeGe_MagSusceptAndOrdering,Bernhard1988_MagneticPhaseDiagramFeGe} and FeSn\cite{Sales2019_2DMagnetismFeSn,Kakihana2019_ElectronicStatesFeSn+CoSn,Yamaguchi1967_FeSnNeutronDiffraction} has been reported and explored elsewhere. The N\'eel transition of FeSn is evident at 365\,K, as is the moment reorientation of FeGe near 100\,K \cite{Beckman1973_LocalVsItinerantInHexFeGe,Beckman1972_hexFeGe_MagSusceptAndOrdering,Bernhard1988_MagneticPhaseDiagramFeGe}. We attribute the clear Curie contribution to the FeGe data below 60\,K to the second phase in the sample.

\subsection{Resistivity}
\label{sec:Results_Resistivity}

\begin{figure}
	\includegraphics{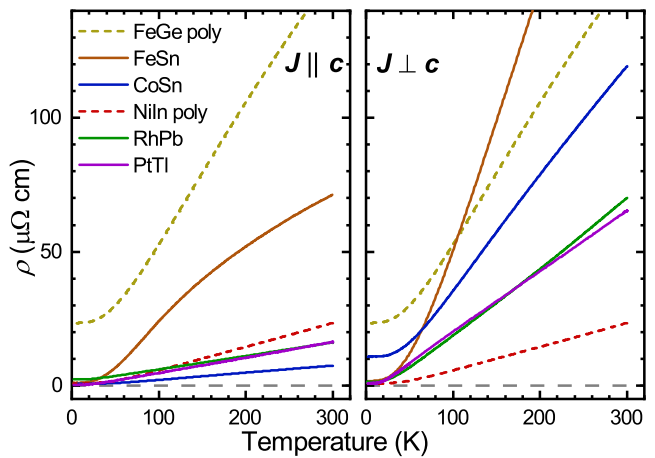}
	\caption{\label{fig:RhoT} Anisotropic electrical resistivity of the kagome metals.}
\end{figure}

The anisotropic resistivity vs temperature, $\rho(T)$, of FeSn, CoSn, RhPb and PtTl are presented in Fig.~\ref{fig:RhoT} along with the resistivity of polycrystalline FeGe and NiIn (dashed). The most obvious common characteristic of these anisotropic data is that the in-plane resistivity rises more quickly with temperature than it does for the $c$-axis. In addition, the residual resistivity values in table~\ref{tab:PropertySummary} are markedly small (order 1\,\textmu$\Omega$\,cm). There are two outliers in this respect; CoSn with $\bm{J}\perp\bm{c}$ and polycrystalline FeGe. In the latter case we believe this large residual resistivity value is also a result of our impure sample and grain boundaries. Finally, we note that there are no discontinuities or changes in slope in any of the $\rho(T)$ curves suggestive of phase transitions.

\subsection{Band structure}
\label{sec:Results_BandStructure}

\begin{figure*}
	\includegraphics{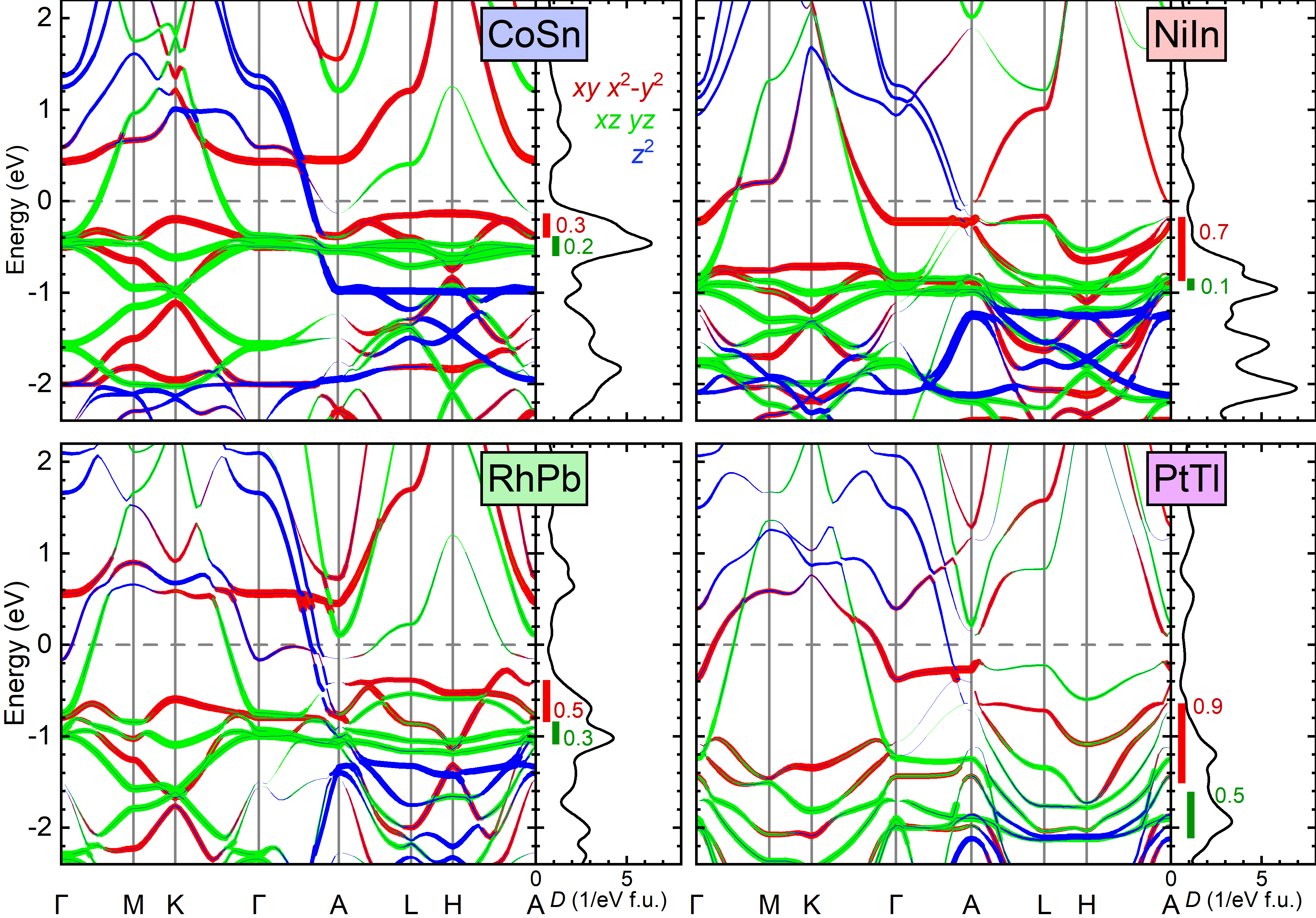}
	\caption{\label{fig:BandStruct+DoS} 
		Band structures of the CoSn-type compounds without magnetic order calculated by DFT. Energy is measured with respect to the Fermi energy. The width of the colors on each band indicate the $d$-orbital character. The density of states, $D$, of each compound is plotted on the right of each panel. The red and green bars mark the approximate extent of the $d_{xy}, d_{x^2-y^2}$ and $d_{xz}, d_{yz}$ flat bands respectively. They are accompanied by the bandwidth in eV.
		%Band structure of the 4 CoSn-type metals without magnetic order calculated by DFT with relaxed lattice parameters. The lower row presents the density of states for each compound and the contributions from each atomic site. The contributions from Sn, In, Pb and Tl are multiplied by 3 to improve their visibility. The Sn1, In1, Pb1, and Tl1 sites are the $1a$ Wyckoff site at the center of the kagome hexagons (see Fig.~\ref{fig:Stucture}).
	}
\end{figure*}

%All atomic sites in the CoSn structure have fixed fractional coordinates in the unit cell. Despite these reduced degrees of freedom, the relaxed lattice parameters are increasingly overestimated from CoSn $\rightarrow$ NiIn $\rightarrow$ RhPb $\rightarrow$ PtTl. Especially for the $c$-axis, which is overestimated by 3.5\% for PtTl (See Appendix~\ref{sec:Appendix_DFTLattice}). 

The DFT band structures calculated for the four compounds with 23 valence electrons are presented in Fig.~\ref{fig:BandStruct+DoS}. The orbital character of each state is represented by the width of the colored lines on each band. The density of states, $D$, of each compound is plotted along the right side of each panel.

In these band structures we can identify replicas of the kagome dispersion (see Fig.~\ref{fig:Stucture} c) for each $d$-orbital character (indicated by color) at the $\Gamma-M-K$ and $A-L-H$ planes (see Fig.~\ref{fig:Stucture}~d). These are most distinct for CoSn, as noted in Refs.~\cite{Kang2020_TopologicalFlatBandsCoSnARPES} and \cite{Liu2020_DiracFermionsFlatBandsCoSn_ARPES}. 

The flat bands nearest the Fermi energy are of greatest interest. We estimate the position and extent of these bands across the Brillouin zone with red and green bars in the density of states plots (for $d_{xy}, d_{x^2-y^2}$ and $d_{xz}, d_{yz}$ flat bands respectively). These generally line up with the peaks in the density of states and are accompanied by the estimated bandwidth (in eV). None of the compounds have flat bands at the Fermi energy. We wish to compare three characteristics of the flat bands: bandwidth, band energy, and distinctness. 

Bandwidth is important as it determines the strength of electron correlations. First we note that the green $d_{xz}, d_{yz}$ flat bands generally are narrower than the red ($d_{xy}, d_{x^2-y^2}$). This is more pronounced for NiIn and PtTl where the $d_{xy}, d_{x^2-y^2}$ band nearest the Fermi energy disperses strongly in the $A-L-H$ plane. In addition, the heavier compounds also show relatively larger bandwidths. These broader bands are also reflected in the small magnitude of the corresponding peaks in the density of states.

To explore flat band physics the bands of interest must be tuned to the Fermi level. Therefore, compounds with flat bands at smaller binding energies will require less doping. CoSn has the nearest flat bands starting at -0.13 and -0.38\,eV. These shift to progressively lower energies in RhPb, NiIn, and PtTl. The band locations are also clearly visible in the position of peaks in the density of states.

Finally, the ``distinctness" of the flat bands is a qualitative measure of how independent they are from other bands. For this purpose, a distinct band is one that maintains its orbital character (color) and small dispersion over most of the Brillouin zone. By this criteria, we would say that the green $d_{xz}, d_{yz}$ band at -0.5\,eV in CoSn is the most distinct. Broadly, the CoSn and NiIn have more distinct flat bands than their heavier counterparts. In PtTl the bands are constantly swapping orbital character and it becomes difficult to identify any features of the kagome lattice dispersion.

%First, we note that the flat bands appear to become less flat for the heavier compounds. In the $3d$ transition metal compounds, CoSn and NiIn, the bandwidth of the low-dispersion bands in 10-200\,meV. In RhPb ($4d$) these band are lest distinct and order 100-400\,meV wide. The flat bands in $5d$ PtTl are broader still and have bandwidths of order 500\,meV. These broader bandwidths are reflected in the reduced magnitude of the peaks in the density of states as well.
%
%\textbf{Why?}
%\textbf{1} 4d and 5d orbitals are more extended giving increased orbital overlap? The kagome flat bands are supposed to originate from hopping interference on lattice. Why would the strength of the hopping matter?
%\textbf{2} Heavier compounds have stronger SOC and therefore band-mixing from it.
%\textbf{3} Other ideas?

%Another key difference between these compounds is the energy of the flat bands relative to the Fermi level. In CoSn these bands are quite close, within a hundred meV. In NiIn and RhPb the flat bands are shifted lower to 700-1000\,meV. And in PtTl they are beyond 1\,eV away.

%The degree of dispersion along the $\mathrm{\Gamma}$-$A$ line also increases from $3d \rightarrow 4d \rightarrow 5d$. This may be caused by increasing coupling between the kagome layers.

%\textbf{?Present Fermi surfaces of CoSn RhPb and PtTl and discuss similarities?}

\section{Discussion}
\label{sec:Discussion}

Our data indicates that CoSn, NiIn, RhPb, and PtTl undergo no phase transitions below room temperature. Bulk superconductivity was not observed in any of the six compounds down to 1.9\,K. %In some samples of RhPb, minor superconductivity of RhPb$_2$ ($T_\mathrm{c} = 2.6$\,K \textbf{Cite SC list}) was observed in resistivity and low-field magnetization. Hints of superconducting Sn and Tl in CoSn and PtTl could be removed by cleaning the crystals ($T_\mathrm{c} = 3.72$ and $2.38$\,K respectively).

In FeGe and FeSn the large density of states from the $3d$-bands lies at the Fermi energy, facilitating robust itinerant antiferromagnetism \cite{Marder_CondensedMatterPhysics}. The other four compounds have an additional valence electron per formula unit, which fills these bands. The reduced density of states at $\varepsilon_\mathrm{Fermi}$ (Fig.~\ref{fig:BandStruct+DoS}) is likely why magnetic order is not observed in these compounds. In fact, antiferromagnetism is systematically suppressed as electron count increases across the (Fe$_{1-x}$Co$_x$)Sn series \cite{Meier2019_AFM_ReorientationCoDopedFeSn,DjegaMariadassou1969_MSn2+MSn}.

The anisotropic resistivity of FeSn, CoSn, RhPb, and PtTl clearly show electronically 3-dimensional metals. We suspect smaller resistivities along the $c$-axes are connected to the strongly dispersing bands passing the Fermi level along the $\Gamma-A$ line in Fig.~\ref{fig:BandStruct+DoS}. 
%\textbf{Discuss how considering the Fermi surfaces and the Boltzmann transport equations give lower resistivity $\bm{J}\parallel\bm{c}$?}

\subsection{Site order in PtTl}
\label{sec:Discussion_PtTl_Order}

PtTl presents a unique challenge for structural refinement. Its two constituents have relatively little difference in atomic number ($_{78}$Pt\,$_{81}$Tl). This creates significant uncertainty in the assignment of elements on the 3 sites in the CoSn structure and Zintl did not suggest a site assignment in the original report \cite{Zintl1935_StructPlatinumThallium}. We suspect that the transition metal, Pt, occupies only the Co-site (3f) forming the kagome lattice as observed in all the other members of this family. 

We have two pieces of evidence for this configuration. The small residual resistivity of PtTl suggests it is an ordered compound. If it was a random alloy (Pt and Tl equally likely on all sites), we estimate a residual resistivity of order 30-100\,\textmu $\Omega$\,cm (using method in Refs.~\cite{Butler1985_TheoryOfElectronicTransportInRandomAlloys,Mu2019_ElectronScatteringMechanismsInNiFeCoCrMnHighEntropyAlloy,Samolyuk2019_OpticalConductivityOfMetalAlloysWithResisdualResistivitiesNearMIRLimit}). This is significantly larger than the measured 0.23 and 0.815\,\textmu $\Omega$\,cm along the $c$-axis and in-plane, respectively.

In addition to this resistivity argument, first principals calculations suggest that PtTl with an inverted CoSn structure (with Tl forming the kagome lattice) has a formation energy of 5.25\,eV/unit cell higher. Anti-site defect pairs (one pair of Pt and Tl atoms swapped) in the standard CoSn structure are also unlikely with formation energies of 2.20\,eV. Therefore, we believe that PtTl is likely well ordered with the CoSn structure and Pt forming the kagome lattice.

\subsection{Susceptibility model}
\label{sec:Discussion_SusceptibilityModel}

\begin{figure}
	\includegraphics{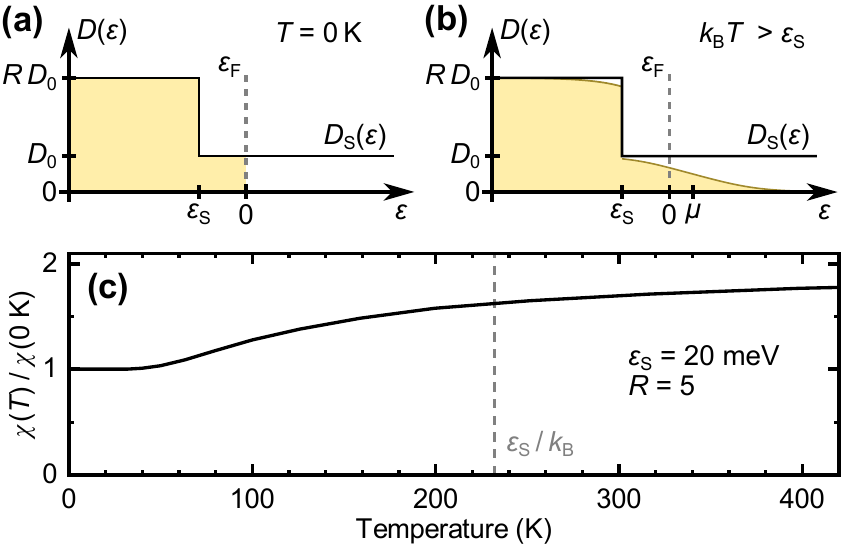}
	\caption{\label{fig:DoSModel} Simple density of states model for the magnetic susceptibility of CoSn. \textbf{a} The step function model for density of states at $T=0$\,K shaded to show the filled states. \textbf{b} When $k_\mathrm{B} T \geq \varepsilon_\mathrm{S} / 4$ states below  $\varepsilon_\mathrm{S}$ begin to thermally depopulate. This allows more electrons to participate in spin polarization and the Pauli susceptibility is enhanced. In addition, the chemical potential $\mu$ shifts. \textbf{c} Plots of the relative change in Pauli susceptibility with temperature based on the density of states model (with $\varepsilon_\mathrm{S}=20$\,meV and $R=5$) including the shifting chemical potential. The dashed line represents the temperature corresponding to the energy scale of $\varepsilon_S$.}
\end{figure}

Lets explore our magnetic susceptibility data for these kagome metals in greater depth as it provides additional insights into the electronic bands structure of these materials. Consider some common contributions to a metal's magnetic response: 
\begin{itemize}
	\item Core diamagnetism
	\item Landau diamagnetism
	\item Pauli paramagnetism
	\item Curie-Weiss 
	\item Ordered moments
\end{itemize}

First, the antiferromagnets FeGe and FeSn are dominated by the last two contributions giving them relatively large, positive susceptibilities (Fig.~\ref{fig:ChiT+Mag}). 

Magnetic order is not evident in the other four compounds so we suspect that the small Curie contributions to the susceptibility at low-temperature (Fig.~\ref{fig:ChiT}) arise from impurity phases or defects in the crystals. Each of these compounds should also have a core diamagnetic contribution we estimate based on the noble gas cores in each compound in Table~\ref{tab:PropertySummary} from a table in  Ref.~\cite{Kittel2004_SolidStatePhysics}.

The most striking feature in Fig.~\ref{fig:ChiT} is the pronounced anisotropy between the $\bm{H}\parallel\bm{c}$ and $\perp\bm{c}$ data. The difference is essentially temperature-independent and the susceptibility along the $c$-axis is always smaller. 

In the absence of ordered moments, we ascribe this anisotropy to Landau diamagnetism. For a spherical Fermi surface, this is expected to be isotropic and proportional to the density of states at the Fermi energy\cite{Kittel2004_SolidStatePhysics,Grosso2014_SolidStatePhysics}. Anisotropy can arise when the totaled magnetization contributions of Landau-orbits on non-spherical Fermi surfaces are strongly dependent on field direction\cite{Nikolaev2018_LandauDiamagMetalFermiSurface}. This had little bearing on our investigation of flat bands so we did not investigate further.

%\textbf{?Discuss commonalities between CoSn RhPb and PtTl? Show fermi surfaces and compare to emphasize point?}.

The temperature dependence of CoSn's susceptibility data is markedly different from the others in Fig.~\ref{fig:ChiT}. It shows a smoothly increasing, S-shaped $\chi(T)$ on heating above 50\,K. Critically, this contribution is basically identical in both directions. 

We attribute this isotropic susceptibility enhancement in CoSn to additional Pauli paramagnetic contributions of kagome flat bands near the Fermi energy. In brief, as temperature increases, electron states farther from the $\varepsilon_\mathrm{Fermi}$ are thermally depopulated and begin contributing to the Pauli susceptibility. In this way, the jump in the density of states from the top of the flat bands near the Fermi energy (see density of states plots in Fig.~\ref{fig:BandStruct+DoS}~a) can cause larger susceptibility at high temperatures.

This makes temperature dependent magnetization measurements a useful tool for screening materials for the peaks and discontinuities in the density of states, $D(\varepsilon)$, near the Fermi energy\cite{Jaccarino1967_ParamagnetismFeSi,Sales1994_MagneticTrasportStruct_FeIrSi,Mandrus1995_ThermodynamicsOfFeSi,Sales2010_SemimetalModelBaFeCo2As2}. Now we will present a model for the density of states to estimate the energy difference between the top of the kagome flat bands and the Fermi energy in CoSn.

The expression for Pauli spin susceptibility can be written \cite{AshcroftMermin1976_SolidStatePhysics,Sales1994_MagneticTrasportStruct_FeIrSi,Mandrus1995_ThermodynamicsOfFeSi},
\begin{equation}
	\label{equ:ChiFromDoS}
	\chi_\mathrm{Pauli} = - \mu_\mathrm{B}^2 \int_{-\infty}^{\infty} \frac{df}{d\varepsilon}(\varepsilon)\,D(\varepsilon)\,d\varepsilon .
\end{equation}
\noindent $\mu_\mathrm{B}$ is the Bohr magneton, $\frac{df}{d\varepsilon}$ is the energy derivative of the Fermi-Dirac distribution, $f(\varepsilon) = 1 / (e^{\frac{\varepsilon-\mu}{k_\mathrm{B}\,T}} + 1)$ and $D(\varepsilon)$ is the density of states including spin degeneracy. $\mu$ is the chemical potential, $k_\mathrm{B}$ is Boltzmann constant and $T$ is temperature.

Looking at the density of states plot for CoSn in Fig.~\ref{fig:BandStruct+DoS}, we approximate the dramatic increase in $D(\varepsilon)$ just below the Fermi energy by a step function depicted in Fig.~\ref{fig:DoSModel}~a;

\begin{equation}
	\label{equ:StepDoS}
	D_\mathrm{S}(\varepsilon) = 
	\left\{
		\begin{array}{ll}
			D_0 & \varepsilon > \varepsilon_S \\
			R\,D_0 & \varepsilon < \varepsilon_S \\
		\end{array}
	\right. ,
\end{equation}

\noindent where $D_0$ is the density of states at the Fermi energy and $R$ is the ratio of the larger density of states to $D_0$.

Fortunately, plugging this into Eq.~\ref{equ:ChiFromDoS} yields an analytical result:
\begin{equation}
	\label{equ:ChiFromStepDoS}
	\chi_\mathrm{Pauli}(T) = \mu_\mathrm{B}^2\,D_0\,(R-1)\,\dfrac{e^{\frac{\varepsilon_\mathrm{S}-\mu}{k_\mathrm{B}\,T}}}{e^{\frac{\varepsilon_\mathrm{S}-\mu}{k_\mathrm{B}\,T}} + 1} + \mu_\mathrm{B}^2\,D_0 .
\end{equation}

Figure~\ref{fig:DoSModel} sketches the origin of the rising $\chi(T)$. At low temperatures (panel a) only the states at the Fermi energy can contribute to Pauli susceptibility. At higher temperatures (Fig.~\ref{fig:DoSModel}~b) states below $\varepsilon_S$ begin to thermally depopulate and the Pauli susceptibility increases as more electrons can participate in the spin polarization in field. These extra excited electrons push the chemical potential, $\mu$, higher (see appendix). 

Figure~\ref{fig:DoSModel}~c presents magnetic susceptibility is enhanced with temperature in this model. This S-shaped curve closely matches that seen in the susceptibility data for CoSn. We can fit susceptibility data with
\begin{equation}
	\label{equ:ChiTerms}
	\chi(T) = \chi_\mathrm{Pauli}(T) + \chi_\mathrm{CW}(T) + \chi_0,
\end{equation}
\noindent which incorporates a Curie term, $\chi_\mathrm{CW}(T) = C/(T-\theta)$, and a temperature-independent contribution, $\chi_0$ (including Landau and core diamagnetism). Unfortunately, multiple temperature-independent contributions mean we can only extract $\varepsilon_S$ from fits to the CoSn $\chi(T)$ data as $\chi_0$, $D_0$, and $R$ are not independent.

Fitting Eq.~\ref{equ:ChiTerms} (with $\mu=0$\,meV) to the CoSn data for $\bm{H}\parallel\bm{c}$ and $\perp\bm{c}$ yielded values of $\varepsilon_\mathrm{S}$ of -19 and -15\,meV, respectively. These values are of the same small magnitude as flat bands centered at -70\,meV in the ARPES results in Ref.~\cite{Liu2020_DiracFermionsFlatBandsCoSn_ARPES}. The susceptibility of CoSn further increases above 400\,K (see appendix~\ref{sec:Appendix_CoSnHighTempChi}). This hints at additional discontinuities in the $D(\varepsilon)$ in the -100 to -200\,meV range.

The $\chi(T)$ data for NiIn, RhPb, and PtTl show far weaker temperature dependence and no inflection point like that expected from Equ.~\ref{equ:ChiFromStepDoS}. In the context of our Pauli susceptibility model, this suggests that either the flat bands are farther from the Fermi energy or the change in the density of states is less dramatic. With $\mu=0$\,meV, the inflection point of Equ.~\ref{equ:ChiFromStepDoS} is expected at $k_\mathrm{B}\,T = 0.42\,|\varepsilon_\mathrm{S}|$. No inflection is observed in these 3 compounds below 300\,K suggesting there are no sharp discontinuities in the density of states within 60\,meV of the Fermi level. This is consistant with the locations of the flat bands in Fig.~\ref{fig:BandStruct+DoS}.

\subsection{Where the flat bands are}
\label{sec:Discussion_FlatBands}

Based on our results, CoSn, NiIn, RhPb, and PtTl do not have flat bands at the Fermi energy. The DFT results in Fig.~\ref{fig:BandStruct+DoS} suggest this situation. A smaller density of states at $\varepsilon_\mathrm{Fermi}$ is supported by the smaller $\gamma$ values measured for these four compounds compared to the two antiferromagnets.

Of the four 23-electron compounds, CoSn appears to have flat bands closest to $\varepsilon_\mathrm{Fermi}$. DFT results in Fig.~\ref{fig:BandStruct+DoS} show the top of the low-dispersion bands about 100\,meV below. CoSn shows an increasing magnetic susceptibility with temperature which we model to arise from a discontinuity in the $D(\varepsilon)$ about 20\,meV below $\varepsilon_\mathrm{Fermi}$. ARPES studies support this interpretation \cite{Liu2020_DiracFermionsFlatBandsCoSn_ARPES,Kang2020_TopologicalFlatBandsCoSnARPES}. 

Our DFT calculations and magnetic susceptibility data suggest the flat bands are more distant in NiIn, RhPb, an PtTl. Their nearly-flat $\chi(T)$ data suggests no abrupt variations in $D(\varepsilon)$ within about $60$\,meV of $\varepsilon_\mathrm{Fermi}$.

Next, lets examine how the 23 electron CoSn-type compounds compare as platforms for future studies of flat band physics. Of these systems, CoSn is the best starting point for exploring flat band physics with its narrow flat band near the $\varepsilon_\mathrm{Fermi}$. Hole doping should drop the Fermi energy into the flat bands. The large $D(\varepsilon_\mathrm{Fermi})$ this creates is likely the origin of antiferromagnetism in the (Fe,Co)Sn series \cite{Meier2019_AFM_ReorientationCoDopedFeSn,DjegaMariadassou1969_MSn2+MSn}. Other doping series might favor different ground states.

NiIn is a less attractive candidate for study. First, it isn't obvious how to obtain single crystals (see section \ref{sec:Methods_growth}). On top of this, the exciting flat bands are deeper in energy and the nearest band is quite broad (see Fig.~\ref{fig:BandStruct+DoS}).

Magnetic order is one way to reduce the large degeneracy associated with flat bands at the Fermi level. This degeneracy likely drives the antiferromagnetic ground states of FeGe, FeSn and (Fe,Co)Sn. Notably, Rh and Pt do not frequently form magnetic compounds. If we succeed in tuning $\varepsilon_\mathrm{Fermi}$ into flat bands composed of Rh $4d$ or Pt $5d$ states, how will these systems reduce this degeneracy? Can we induce a more interesting ground state such as a structural distortion, charge density waves, or superconductivity? 

RhPb deserves some additional attention despite its expensive reactant and flat bands lying below -400\,meV. In Fig.~\ref{fig:BandStruct+DoS} these bands still have relatively small bandwidths. We can estimate the doping required to tune these bands to the Fermi energy by integrating the density of states from $\varepsilon_\mathrm{Fermi}$ down to the top of the flat bands (-400\,meV). This suggests we need to dope RhPb by 0.4\,holes/formula unit to bring these bands to $\varepsilon_\mathrm{Fermi}$. In addition to the potential for unusual ground states, RhPb is a heavier analog of CoSn and will allow us to explore the impact of stronger spin-orbit coupling on flat band physics.

PtTl is the least attractive candidate. The kagome lattice flat bands are even broader and lower in energy than RhPb. It not only includes expensive Pt but also toxic and reactive Tl. On top of this, we estimate at least 0.8\,holes/formula unit would be required to raise the flat bands about 1\,eV to the Fermi energy.

Now, lets generalize what we have learned here to other systems where flat bands are generated by geometric frustration. $3d$ transition metals tend to give narrower bandwidths for these bands than $4d$ and $5d$ versions, but the heavier atoms offer stronger spin orbit coupling. Next, CoSn and NiIn are entirely iso-electronic but NiIn has broader flat bands farther from $\varepsilon_\mathrm{Fermi}$. This means that more favorable flat band energies might be obtained by exploring variations of existing flat band systems with different element pairs.

\section{Conclusions}
\label{sec:Conclusions}

Kagome lattice materials can host flat bands resulting from geometric frustration in their structure. The CoSn-type materials host stacked kagome lattices of transition-metals atoms. 

In this paper we explore the six compounds with the CoSn structure type. We make samples of each CoSn-type compound including single crystals of FeSn, CoSn, RhPb, and PtTl. Curiously, we find CoSn and NiIn are subtly colored metals.

We report the low-temperature thermal expansion, heat capacity, magnetic susceptibility, and resistivity of these compounds. FeGe and FeSn are antiferromagnets, but we do not observe any phase transitions in CoSn, NiIn, RhPb and PtTl between 1.9 and 380\,K. The single crystals of these metals exhibit both resistive and pronounced magnetic anisotropies. 

DFT band structure calculations reveal the location and character of flat bands in the CoSn-type materials without magnetic order. Critically, these calculations and the experimental data suggest that none of these have flat bands at the Fermi level. Of the four, CoSn has flat bands nearest the Fermi energy. We also note that the flat bands have increasing bandwidth for $3d\rightarrow4d\rightarrow5d$ transition-metals.

CoSn shows an increasing magnetic susceptibility on heating above 50\,K. We model this with additional Pauli paramagnetic contributions by thermal depopulation of flat bands about 20\,meV below the Fermi energy. The magnetic data of NiIn, RhPb, and PtTl show far weaker variations with temperature. This suggests flat bands are more than 60\,meV below $\varepsilon_\mathrm{Fermi}$ in these cases, in agreement with our DFT calculations.

We favor CoSn for future investigations of flat band physics because its flat bands are narrow and near the Fermi energy. RhPb also deserves attention because of its heavier constituents and relatively-distinct flat bands. This makes it a good system for investigating flat band physics with $4d$ orbitals and stronger spin orbit coupling. NiIn and PtTl are less attractive starting points.

The CoSn family of kagome metals provides a nice playground to explore physics in lattice-derived flat bands. The diverse family members include $3d$, $4d$, and $5d$ elements facilitating investigation of the role of $d$-orbital character and spin orbit coupling strength.

\begin{acknowledgments}
	\label{sec:Acknowledgment}
	The authors thank Anna B\"ohmer, Dmitry Chichinadze, Anchal Padukone, and Jiaqiang Yan for their helpful discussions and insights. We would also like to acknowledge that NiIn is a palindrome.
		
	Research supported by the U. S. Department of Energy, Office of Science, Basic Energy Sciences, Materials Sciences and Engineering Division (under contract number DE-AC05-00OR22725). High-temperature X-ray diffraction experiments (C.A.B.) were sponsored by the Laboratory Directed Research and Development program of Oak Ridge National Laboratory, managed by UT-Battelle, LLC, for the U.S. Department of Energy. GDS was supported as part of the Energy Dissipation to Defect Evolution (EDDE), an Energy Frontier Research Center funded by the US Department of Energy, Office of Science, Basic Energy Sciences under Contract Number DE-AC05-00OR22725.
\end{acknowledgments}

%\bibliography{CoSn-Type_Kagome_Metals_10_sub.bbl}% Produces the bibliography via BibTeX.

%

\appendix

\section{High temperature CoSn susceptibility}
\label{sec:Appendix_CoSnHighTempChi}

\begin{figure}[h]
	\includegraphics{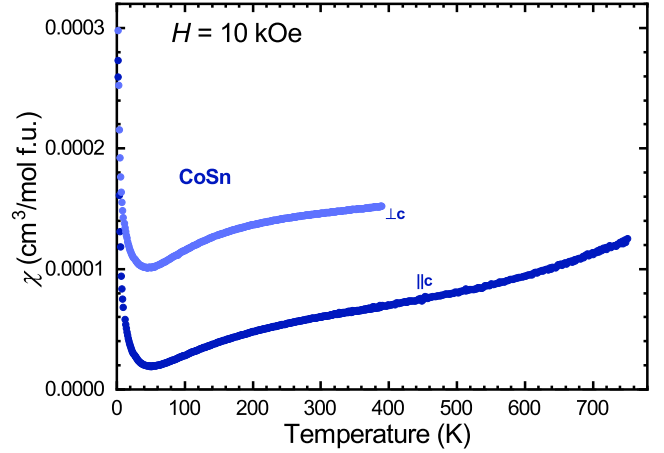}
	\caption{\label{fig:ChiT_CoSnHighT} Anisotropic magnetic susceptibility of the CoSn up to 750\,K.}
\end{figure}

In section~\ref{sec:Discussion_SusceptibilityModel} we model CoSn's rising magnetic susceptibility as a temperature dependent Pauli paramagnetic response. Figure~\ref{fig:ChiT_CoSnHighT} shows how the magnetic susceptibility of CoSn continues to increase above 400\,K. In our model, this suggests further discontinuities in the density of states 100-200\,meV below $\varepsilon_\mathrm{Fermi}$.

%\section{DFT lattice parameters}
%\label{sec:Appendix_DFTLattice}
%
%\begin{table}
%	\caption{Lattice parameters of CoSn-type compounds at 15\,K compared to those of the relaxed compounds in DFT.
%		\label{tab:DFTLattice}}
%	\begin{tabular}{l|c|c|c||c|c|c}
%		 &$a$(15\,K) &$a_\mathrm{DFT}$ &error &$c$(15\,K) &$c_\mathrm{DFT}$ &error\\
%		&(\AA) &(\AA) & &(\AA) &(\AA) &\\
%		\hline
%		CoSn &5.26926(6) &5.290 &0.39\% &4.24310(5) &4.224 &-0.45\%\\
%		NiIn &5.22955(9) &5.276 &0.89\% &4.33892(7) &4.376 &0.85\%\\
%		RhPb &5.66601(20) &5.731 &1.15\% &4.41267(10) &4.505 &2.09\%\\
%		PtTl &5.60172(5) &5.698 &1.72\% &4.62755(3) &4.787 &3.45\%\\
%		
%	\end{tabular}
%\end{table}
%
%Table~\ref{tab:DFTLattice} shows that the relaxed lattice parameters from DFT show increasing error for the heavier CoSn-type compounds.

\section{Pauli Susceptibility model}
\label{sec:AppendixSusceptModel}

For our simple $D(\varepsilon)$ model (Equ.~\ref{equ:StepDoS}) we can derive an analytical expression for the change in chemical potential with temperature (setting $\mu=0$ at $T=0$). When the density of states step is below the Fermi level ($\varepsilon_\mathrm{S}<0$) then the chemical potential $\mu$ is the solution to;
\begin{equation}
\label{equ:mu_neg_eStep}
\mu = (R-1)\,k_\mathrm{B}\,T\, \ln\left(e^{\frac{\varepsilon_\mathrm{S}-\mu}{k_\mathrm{B}\,T}} + 1\right)
%	e^{\frac{1}{R-1}\,\frac{\mu}{k_\mathrm{B}T}} = e^{\frac{\varepsilon_\mathrm{S}-\mu}{k_\mathrm{B}T}} + 1.
\end{equation}
\noindent When the step is above, the chemical potential $\mu$ is the solution to;
\begin{equation}
	\label{equ:mu_pos_eStep}
	\mu=(R-1)\,\left(-\varepsilon_\textrm{S}+\,k_\mathrm{B}\,T\, \ln\left(e^{\frac{\varepsilon_\mathrm{S}-\mu}{k_\mathrm{B}\,T}} + 1\right)\right)
%	e^{\frac{1}{R-1}\,\frac{\mu}{k_\mathrm{B}T}}\,e^{\frac{\varepsilon_\mathrm{S}}{{k_\mathrm{B}T}}}= e^{\frac{\varepsilon_\mathrm{S}-\mu}{k_\mathrm{B}T}} + 1.
\end{equation}

The magnetic susceptibility (Equ.~\ref{equ:ChiFromStepDoS}) and chemical potential (Equ.~\ref{equ:mu_neg_eStep}) both begin to increase at the same temperature. As a result, the inflection point in Fig.~\ref{fig:DoSModel}~c is a good measure of the $\varepsilon_S$.

\end{document}